\documentclass[sigconf]{acmart}
\usepackage[utf8]{inputenc}
\usepackage{lipsum}     
\usepackage{tabularx}   
\usepackage{array}      
\usepackage{listings}
\usepackage{xcolor} 
\lstdefinelanguage{MyARM}{
    morekeywords={ldr, str, mov, add, sub, b, bl, br, cbz, adr}, 
    sensitive=false, 
    morecomment=[l]{//}, 
    morecomment=[l]{@}  
}
\usepackage{amsmath}   
\usepackage{algorithm}
\usepackage[noend]{algpseudocode} 
\algrenewcommand\algorithmicrequire{\textbf{Input:}}
\algrenewcommand\algorithmicensure{\textbf{Output:}}
\algnewcommand{\Gets}{\gets} 

\usepackage{float}
\usepackage{url}
\usepackage{hyperref}

\usepackage{tabularx}  
\usepackage{makecell}  
\usepackage{threeparttable} 
\usepackage{array} 
\usepackage{siunitx}

\usepackage{hyperref} 
\usepackage{xcolor}
\usepackage{enumitem}

\AtBeginDocument{%
  }

\acmISBN{978-1-4503-XXXX-X/2018/06}

\begin{document}

\title{XTrace: A Non-Invasive Dynamic Tracing Framework for Android Applications in Production}

\author{Qi Hu}
\affiliation{%
  \institution{ByteDance}
  \country{China}
}
\email{huqi.steven@bytedance.com}

\author{JiangChao Liu}
\affiliation{%
  \institution{ByteDance}
  \country{China}
}
\email{liujiangchao@bytedance.com}

\author{Xin Yu}
\affiliation{%
  \institution{ByteDance}
  \country{Singapore}
}
\email{yuxin@bytedance.com}

\author{Lin Zhang}
\affiliation{%
  \institution{ByteDance}
  \country{China}
}
\email{zhanglin1@bytedance.com}

\author{Edward Jiang}
\affiliation{%
 \institution{TikTok Inc}
 \country{USA}
}
\email{edward.jiang@bytedance.com}

\renewcommand{\shortauthors}{Qi Hu et al.}
\begin{abstract}
As the complexity of mobile applications grows exponentially and the fragmentation of user device environments intensifies, ensuring online application stability faces unprecedented challenges. Traditional methods, such as static logging and post-crash analysis, lack real-time contextual information, rendering them ineffective against "ghost bugs" that only manifest in specific scenarios. This highlights an urgent need for dynamic runtime observability: intercepting and tracing arbitrary methods in production without requiring an app release.
 We propose XTrace, a novel dynamic tracing framework. XTrace introduces a new paradigm of non-invasive proxying, which avoids direct modification of the virtual machine's underlying data structures. It achieves high-performance method interception by leveraging and optimizing the highly stable, built-in instrumentation mechanism of the Android ART virtual machine. 
Evaluated in a ByteDance application with hundreds of millions of daily active users, XTrace demonstrated production-grade stability and performance. Large-scale online A/B experiments confirmed its stability, showing no statistically significant impact (p > 0.05) on Crash User Rate or ANR rate, while maintaining minimal overhead (<7 ms startup latency, <0.01 ms per-method call) and broad compatibility (Android 5.0–15+). Critically, XTrace diagnosed over 11 severe online crashes and multiple performance bottlenecks, improving root‑cause localization efficiency by over 90\%. This confirms XTrace provides a production-grade solution that reconciles the long-standing conflict between stability and comprehensive coverage in Android dynamic tracing.
\end{abstract}




\maketitle

\section{Introduction}
\subsection{Research Background}
Android applications have become increasingly integral to everyday life. Consequently, their complexity is ever-expanding. These applications have evolved into sophisticated systems integrating numerous third-party SDKs, complex business logic, multi-threaded concurrent processing, and frequent network interactions. This intrinsic complexity, compounded by the vast fragmentation of user devices across different models, operating system versions, and network environments, makes ensuring the reliability of online applications a formidable challenge. This is particularly true for online "ghost bugs"—elusive issues that only manifest on the specific users, in the specific scenarios, and at the specific time. When such problems occur in a production environment, developers often face the following predicaments:
\begin{itemize}
\item \textit{Logging "black holes"}: Pre-instrumented static logs provide limited information and cannot cover all exceptional execution paths. When an issue arises, the absence of critical contextual information makes it impossible to determine the root cause.
\item  \textit{Difficulty in reproduction}: Online issues are highly dependent on the environmental state at the moment of occurrence, making them often hard to reproduce in a developer's testing environment. Consequently, troubleshooting becomes akin to finding a needle in a haystack.
\item  \textit{Prolonged fix cycles}: The traditional "hypothesize → add logs → release new version → verify" cycle can extend the time required to fix a single online issue to several days or even weeks, severely impacting user experience and product reputation.
\end{itemize}
Against this backdrop, the industry urgently requires a technique that can bestow comprehensive runtime observability upon online applications without necessitating a new release. Such a technique must support the dynamic diagnosis and on-the-fly probing of arbitrary target methods to acquire their complete call stacks, capture contextual state (including input parameters and return values), and even profile their execution performance in real-time. In summary, such a technique enables dynamic interception and tracing of methods in Android applications in production. The objective of this paper is to present such a technique.

\subsection{Conventional Approaches}
Currently, technical solutions for implementing dynamic interception of arbitrary methods on the Android platform can be categorized into two distinct paradigms: one involves static injection via bytecode instrumentation at compile-time, while the other achieves interception through dynamic hooking at runtime.

The former approach typically utilizes bytecode manipulation frameworks, such as ASM~\cite{ASM}\cite{bytecode_instrumentation_survey}, in conjunction with the Android Gradle Plugin's~\cite{AndroidGradlePlugin} transformer API to weave predefined interception logic into the entry points of the target methods during the compilation phase. In production, the interception logic can be triggered as needed. The advantage of this approach lies in its relatively high stability; however, its limitations are equally pronounced. First, large-scale instrumentation significantly inflates the application's package size and slows down compilation speed, making full coverage infeasible in practical engineering and thus limiting method coverage. Second, compile-time instrumentation can hardly be applied to system methods within the Android framework layer (as system class bytecode cannot be modified). Third, any changes to the tracing logic necessitate recompiling and releasing the application, lacking the dynamism required to respond to emergent online issues.

Conversely, runtime dynamic hooking solutions, represented by frameworks like Frida~\cite{Frida}\cite{frida_usage_example}, operate by attaching to the target process at runtime. They directly modify internal virtual machine data structures, such as ArtMethod, or function pointers in memory, inserting trampoline code at the beginning of a function's instruction segment to forcibly alter the original code execution path. The advantage of this method lies in its powerful coverage capabilities, allowing it to intercept any method within the process's memory space, including system-level methods, theoretically achieving nearly 100\% method coverage. However, its invasive implementation introduces fundamental vulnerabilities. Due to its heavy reliance on modifying non-public and volatile internal VM structures, this technique carries inherent compatibility and stability risks across different Android versions and device manufacturer ROMs, making it highly prone to application crashes in complex production environments. This, in turn, leads to high and continuous adaptation and maintenance costs.

In summary, the current technological landscape forces a difficult trade-off: static injection at compile time ensures stability at the cost of coverage and flexibility, while runtime dynamic hooking achieves broad coverage but sacrifices stability and compatibility.
This gap highlights a critical need in production environments: no existing solution delivers stability, comprehensive coverage, and dynamism simultaneously.
In this paper, we propose \textit{XTrace}, a dynamic tracing framework for Android applications, aiming to address theses challenges.

\subsection{Our Contribution}
Dynamic tracing on production Android systems has long been constrained by the trade-off between stability, performance, and tracing coverage. XTrace addresses this with a non-invasive architecture. To ensure stability, it avoids risky in-memory code patching by leveraging the system's native, stable tracing interfaces.
However, these native interfaces are too slow for production use. To overcome this performance bottleneck, XTrace integrates two key optimization mechanisms.

 
 First, it modifies the system's inherent global instrumentation mode to intervenes only on target methods at runtime, drastically reducing runtime overhead. Second, it abandons the system's indiscriminate use of interpreter-based stubs, replacing it with an "adaptive execution mode" mechanism. This mechanism intelligently applies the optimal tracing mode for each method based on its real-time compilation state, thereby preventing performance regression for compiled methods.


Our main contributions can be summarized as follows:
\begin{itemize}
    \item We propose and validate a new non-invasive proxying dynamic tracing paradigm that achieves both stability and high performance.
    \item We design and implement XTrace, a novel industrial-grade dynamic tracing framework based on this paradigm, which is practically useful for the production environment.
    \item We demonstrate the effectiveness of our framework in solving real-world, complex online issues through large-scale deployment in applications with hundreds of millions of daily active users.
\end{itemize}

\section{Overview}
\label{sec:overview}
To show the effectiveness of the XTrace framework in a real-world, complex production environment, this chapter presents a typical online case where a critical issue was successfully located and resolved with XTrace.
\subsection{A Typical "Ghost Bug" Case}
This case involves a real-world crash issue from a ByteDance application, affecting over 40,000 users daily and exhibiting the typical characteristics of a "ghost bug." 

Its crash stack trace is shown in Figure~\ref{fig:crash_stack}. We can see that the entire call stack (from \textit{getDisplay} at the top, through \textit{onDisplayFeaturesChanged}, down to the \textit{Handler} call at the bottom) consists of only system Framework methods. The absence of application-level code
rendered conventional stack trace analysis ineffective.

\begin{figure*}[h!]
\begin{lstlisting}[
    language=Java,
    frame=single,
    basicstyle=\ttfamily\scriptsize
]
java.lang.UnsupportedOperationException: Tried to obtain display from a Context not associated with one. Only visual Contexts (such as Activity or one created with Context#createWindowContext) are associated with displays ...
   at android.app.ContextImpl.getDisplay(ContextImpl.java:3166)
   at android.content.ContextWrapper.getDisplay(ContextWrapper.java:1209)
   // ... (omitted intermediate framework calls) ...
   at androidx.window.extensions.layout.WindowLayoutComponentImpl.onDisplayFeaturesChanged(WindowLayoutComponentImpl.java:250)
   // ... (omitted intermediate framework calls) ...
   at android.os.Handler.dispatchMessage(Handler.java:99)
\end{lstlisting}
\caption{Ghost Crash Stacks}
\label{fig:crash_stack} 
\end{figure*}

\subsection{Problem Analysis and Diagnostic Challenges}
\textbf{Problem Analysis: }The stack trace indicates that the crash occurs when \textit{ContextWrapper\#getDisplay} is ultimately called from \textit{onDisplayFeaturesChanged}. The crash message \textit{"Only visual Contexts... are associated with displays"} reveals that the crash is caused by the current \textit{Context} not being a UI-related \textit{Context}. This implies that the \textit{Context} instance being used is invalid. The key, therefore, is to identify which part of the application-level code supplied this invalid \textit{Context}.

To do so, we first examined the system source code of the \textit{onDisplayFeaturesChanged} method, found in the crash stack trace. As shown in Figure~\ref{fig:addWindowLayoutCode}, The class \textit{WindowLayoutComponentImpl} contains a variable named \textit{mWindowLayoutChangeListeners} for storing \textit{Context} objects. Inside the \textit{onDisplayFeaturesChanged} method, a \textit{get} operation retrieves the \textit{Consumer} associated with the \textit{Context}, which then executes the subsequent logic. The \textit{put} operation for this variable is performed via the \textit{addWindowLayoutInfoListener} method. Therefore, to determine the source of the invalid \textit{Context}, it is necessary to find out what code called the \textit{WindowLayoutComponentImpl\#addWindowLayoutInfoListener} method prior to the crash.
\begin{figure}[h!]
\begin{lstlisting}[
    language=Java,
    frame=single,
    basicstyle=\ttfamily\footnotesize
]
// androidx.window...WindowLayoutComponentImpl.java

Map<...> mWindowLayoutChangeListeners = new ArrayMap<>();
// In `onDisplayFeaturesChanged`, the `Consumer` is derived from the listener's associated `Context`.
void onDisplayFeaturesChanged(...) {
    ...
    for (Context context : getContextsListeningForLayoutChanges()) {
        Consumer<...> layoutConsumer = mWindowLayoutChangeListeners.get(context);
        ...
    }
    ...
}
// The `addWindowLayoutInfoListener` method stores this `Context`
void addWindowLayoutInfoListener(Context context, ...) {
    ...
    mWindowLayoutChangeListeners.put(context, ...)};
    ...
}
\end{lstlisting}
\caption{Code Analysis that Motivated Tracing `addWindowLayoutInfoListener` for the Crash}
\label{fig:addWindowLayoutCode} 
\end{figure}

\textbf{Diagnostic Challenge:} After the aforementioned analysis, the core diagnostic challenge was transformed into a specific technical question: How can we determine which segment of business code called the system framework's \textit{addWindowLayoutInfoListener} method and passed it an invalid \textit{Context} instance before the crash occurred? Since this method is located in the system framework layer, adding logs by modifying the system source code was not an option, even with a new release. Furthermore, standard compile-time bytecode instrumentation is powerless against framework methods, and Frida-like dynamic hooking tools lack the requisite compatibility and stability for such a large-scale production environment.

\subsection{Applying XTrace for Root Cause Analysis}
To resolve this diagnostic dilemma, we applied our XTrace framework. By dynamically deploying a configuration from the cloud, we instructed XTrace to intercept the target method at runtime. Specifically, we configured it to intercept the \textit{addWindowLayoutInfoListener} method within the \textit{WindowLayoutComponentImpl} class. We also configured an action to capture the stack trace, causing it to record and print the caller's complete stack trace whenever the method was invoked.

This successfully captured the complete call chain leading up to the crash. The captured stack trace as shown in Figure~\ref{fig:xtrace_captured_stack}, definitively identified the application level origin of the crash.

\begin{figure*}
\begin{lstlisting}[
    language=Java,
    frame=single,
    basicstyle=\ttfamily\footnotesize
]
at com.bytedance.xtrace.XTrace.intercept(SourceFile:50331679)
at androidx.window.extensions.layout.WindowLayoutComponentImpl.addWindowLayoutInfoListener(WindowLayoutComponer
// ... (intermediate calls within Chromium WebView) ...
at org.chromium.content.browser.device_posture.DevicePosturePlatformProviderAndroid.startListening(chromium-Trichrome
at com.android.webview.chromium.WebViewChromium.evaluateJavascript(chromium-TrichromeWebViewGoogle6432.aab-sta
at android.webkit.WebView.evaluateJavascript(WebView.java:897)
at com.bytedance.lynx.hybrid.webkit.WebKitView.sendEventByJson(SourceFile:33554470)
at com.bytedance.hybrid.spark.page.SparkFragment.onCreateView(SourceFile:50331701)
// ...
at com.bytedance.hybrid.spark.page.SparkActivity.onStart(SourceFile:24)
\end{lstlisting}
\caption{An Application-level Stack Trace Captured by XTrace}
\label{fig:xtrace_captured_stack} 
\end{figure*}

\subsection{Problem Resolution and Effect Verification}
The caller stack trace in Figure~\ref{fig:xtrace_captured_stack}, printed by XTrace after intercepting the \textit{addWindowLayoutInfoListener} method, shows that the problem originated in the \textit{SparkFragment} module. The specific call path began in the \textit{onCreateView} method and, through a series of calls involving \textit{WebView.evaluateJavascript}, ultimately registered an incorrect \textit{Context} with the system service via \textit{addWindowLayoutInfoListener}.

Armed with this precise diagnostic information, the root cause was swiftly addressed, and the fix completely eradicated the crash upon deployment, ultimately reducing the hourly crash rate to zero.

In Summary, through the XTrace framework, without requiring a new application release, we successfully located and resolved an online crash that had plagued the team for weeks, all in under three hours. This demonstrates the framework's capability and practical value in accurately and efficiently localizing complex, cross-layer, and non-deterministic issues in a real-world production environment.
\section{Methodology}
This chapter elaborates on the design and implementation of the XTrace framework. First, we analyze the ART virtual machine's instrumentation mechanism, which serves as the  foundation for this study. Second, we dissect the application paradigm of the native system trace framework and its inherent defects to demonstrate the necessity of a new solution. Finally, we detail the core innovative strategies designed within the XTrace framework to address these deficiencies.
\subsection{The ART Virtual Machine's Instrumentation Mechanism}
 This research starts with an in-depth investigation into the instrumentation~\cite{AndroidInstrumentation} mechanism within the ART~\cite{AndroidART}\cite{art_perf_analysis} virtual machine. This mechanism, which is a core component designed by ART for advanced functionalities such as debugging and performance profiling, has long been under-explored and under-utilized. It provides a crucial foundation for achieving efficient, low-intrusion dynamic analysis. Before elaborating on our proposed solution, it is necessary to systematically explain the technical principles of this mechanism.  

Through an in-depth examination of the ART virtual machine's source code, tracing the process from class loading to method execution, we identified a event listening and dispatching mechanism embedded within the native interpreter's critical execution path.  The interpreter’s execution flow, which every method must pass through during execution, incorporates the predefined monitoring method in Figure~\ref{fig:interpreter_execute_code}. Before any method begins its actual execution, the system checks whether any \textit{InstrumentationListener} instances are registered. If at least one such listener exists, a \textit{MethodEnterEvent} is triggered, temporarily diverting control to the instrumentation framework. This design confirms that ART intrinsically supports a stable and reliable interception mechanism, enabling observation at key stages of method execution, and thereby establishes the theoretical foundation for our dynamic tracing approach.

\begin{figure}[h!]
\begin{lstlisting}[
    language=C++,
    frame=single,
    basicstyle=\ttfamily\footnotesize
]
// art/runtime/interpreter/interpreter.cc
static inline JValue Execute(Thread* self, /*...*/) {
...
 if (instrumentation->HasMethodEntryListeners()) {
      instrumentation->MethodEnterEvent(self, ...);
      ...
 }
...
}
\end{lstlisting}
\caption{The Instrumentation Checkpoint in \textit{Execute}}
\label{fig:interpreter_execute_code} 
\end{figure}

The operational logic of this mechanism is structured around a robust Publish–Subscribe pattern, which revolves primarily around the \textit{InstrumentationListener} interface. The \textit{InstrumentationListener} serves as a pure virtual base class, defining a standardized suite of callback interfaces through which notifications—generated by the virtual machine during various runtime events (such as \textit{MethodEntered} and \textit{MethodExited}) are received. This design furnishes a comprehensive set of capabilities essential for implementing sophisticated dynamic analysis tasks. External modules, acting as subscribers, utilize the clearly defined registration interface \textit{Instrumentation::AddListener} to enroll an instance of an \textit{InstrumentationListener} implementation, along with its specified event types, into the virtual machine’s internal listener registry. Upon the occurrence of a particular runtime event (i.e., the publishing action) the virtual machine traverses the relevant listener list and systematically invokes the corresponding callback method of each registered listener, thereby accomplishing event dispatch.

The ART instrumentation mechanism is highly stable, because it originates from native implementation. This makes it an ideal technical cornerstone for building a high-stability dynamic tracing framework in the production environment.

\subsection{Defects of the Native Trace Framework}
Although the ART instrumentation mechanism provides a theoretical foundation for dynamic tracing, its application paradigm within the Android system's native trace framework suffers from fundamental design flaws that lead to severe performance issue. This section will dissect the implementation of this native framework to reveal its inherent limitations.

The method tracing functionality in the Android system is typically initiated via the \textit{Debug.startMethodTracing }interface, the call chain of which ultimately executes the Trace::Start~\cite{AndroidTrace} method in the ART virtual machine.  As show in Figure~\ref{fig:trace_start_code}, this method has two core actions: first, it registers a Trace instance as a listener via \textit{Instrumentation::AddListener}; second, it globally activates tracing by calling \textit{Instrumentation::EnableMethodTracing}.

\begin{figure}[h!]
\begin{lstlisting}[
    language=C++,
    frame=single,
    basicstyle=\ttfamily\footnotesize
]
void Trace::Start(...) {
  ...
  the_trace_ = new Trace(...);
  runtime->GetInstrumentation()->AddListener(the_trace_, ...);
  runtime->GetInstrumentation()->EnableMethodTracing(...);
  ...
}
\end{lstlisting}
\caption{The Native Tracing Activation Point in \textit{Trace::Start}}
\label{fig:trace_start_code} 
\end{figure}

\begin{figure}[h!]
    \centering
    \includegraphics[width=1\linewidth]{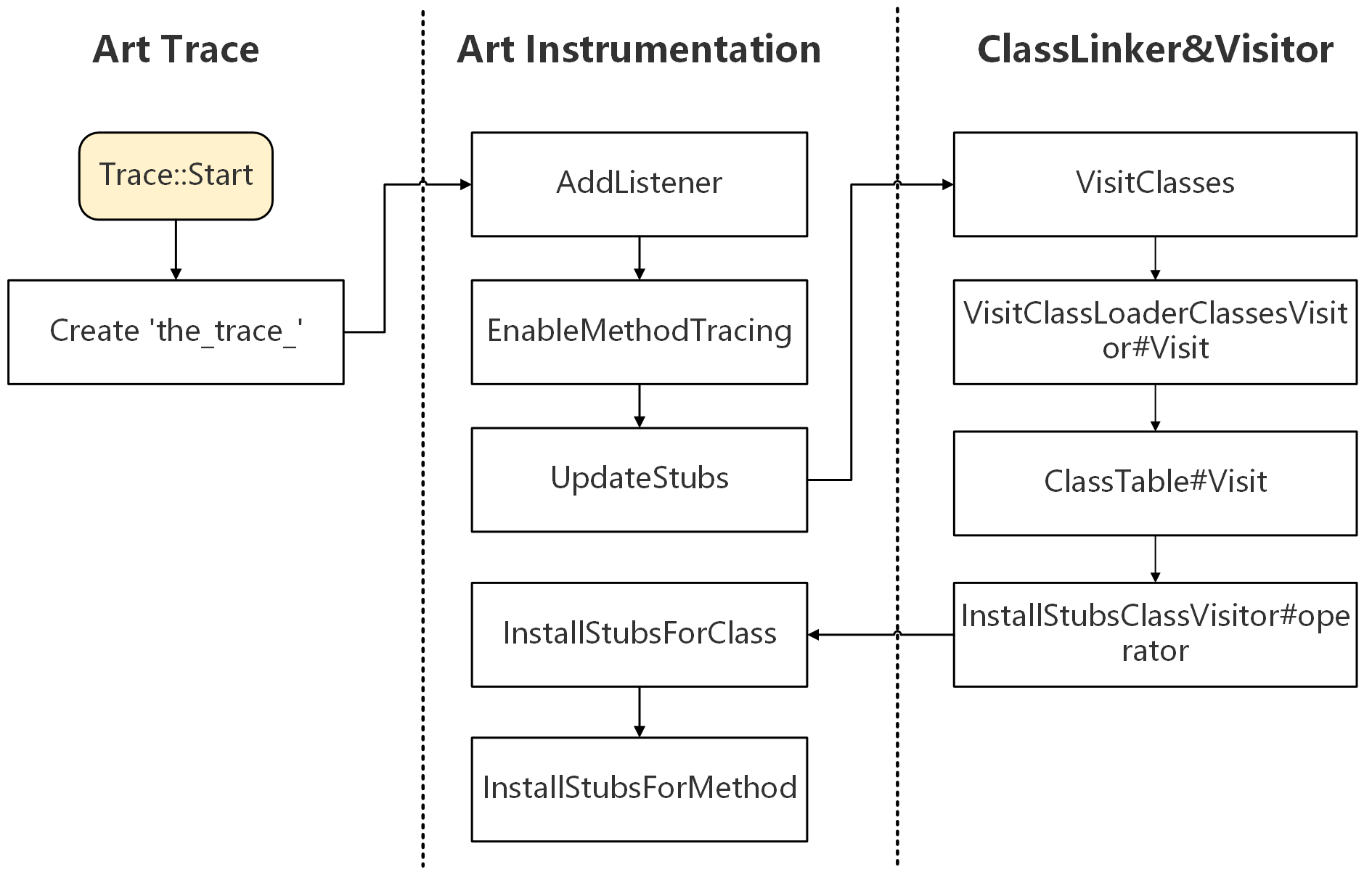}
    \caption{Workflow of the Global Stub Installation}
    \label{fig:system_trace}
\end{figure}

The root of the performance issue stems from the implementation of the \textit{EnableMethodTracing}. As illustrated in Figure~\ref{fig:system_trace}, this method initiates an indiscriminate and system-wide replacement of method entry points. It iterates through all currently loaded classes in the virtual machine via \textit{ClassLinker::VisitClasses} and modifies the entry point for every method of every class (\textit{ArtMethod}).

This static full-coverage strategy is the primary reason for its poor performance, as it imposes unnecessary overhead on a vast number of methods unrelated to the tracing target.

More severely, the framework defaults to a strategy of forced execution mode degradation. That is,  
all native machine code, which should have been executed efficiently after being optimized by Just-In-Time (JIT) or Ahead-Of-Time (AOT) compilation, is forcibly degraded to the least efficient interpreter execution mode. This strategy is implemented by replacing the entry points of all methods with a stub that points to the interpreter bridge in \textit{Instrumentation::InstallStubsForMethod} from Figure~\ref{fig:system_trace}. 
This strategy entirely negates the virtual machine's compilation optimizations, leading directly to a severe degradation in application performance.

In conclusion, while the design paradigm of the native system trace framework remains stable due to its reliance on the instrumentation mechanism, its two fundamental flaws (namely, indiscriminate global injection and the forced degradation to interpreter execution) collectively lead to significant performance degradation. As a result, the framework is unsuitable for performance-sensitive production environments. This analysis underscores that constructing a high-performance dynamic tracing framework necessitates the development of a fundamentally new application paradigm for instrumentation, one capable of circumventing the two aforementioned shortcomings.

\subsection{The Design and Implementation of XTrace}
\subsubsection{Core Idea}
To address the aforementioned defects, we designed and implemented the XTrace framework. The core idea of XTrace is to circumvent the performance deficiencies of the native trace framework by designing a new application paradigm centered on \textit{targeted injection} and \textit{adaptive stub redirection}, all while fully leveraging the inherent stability of the native instrumentation mechanism.

\begin{figure}
    \centering
    \includegraphics[width=1\linewidth]{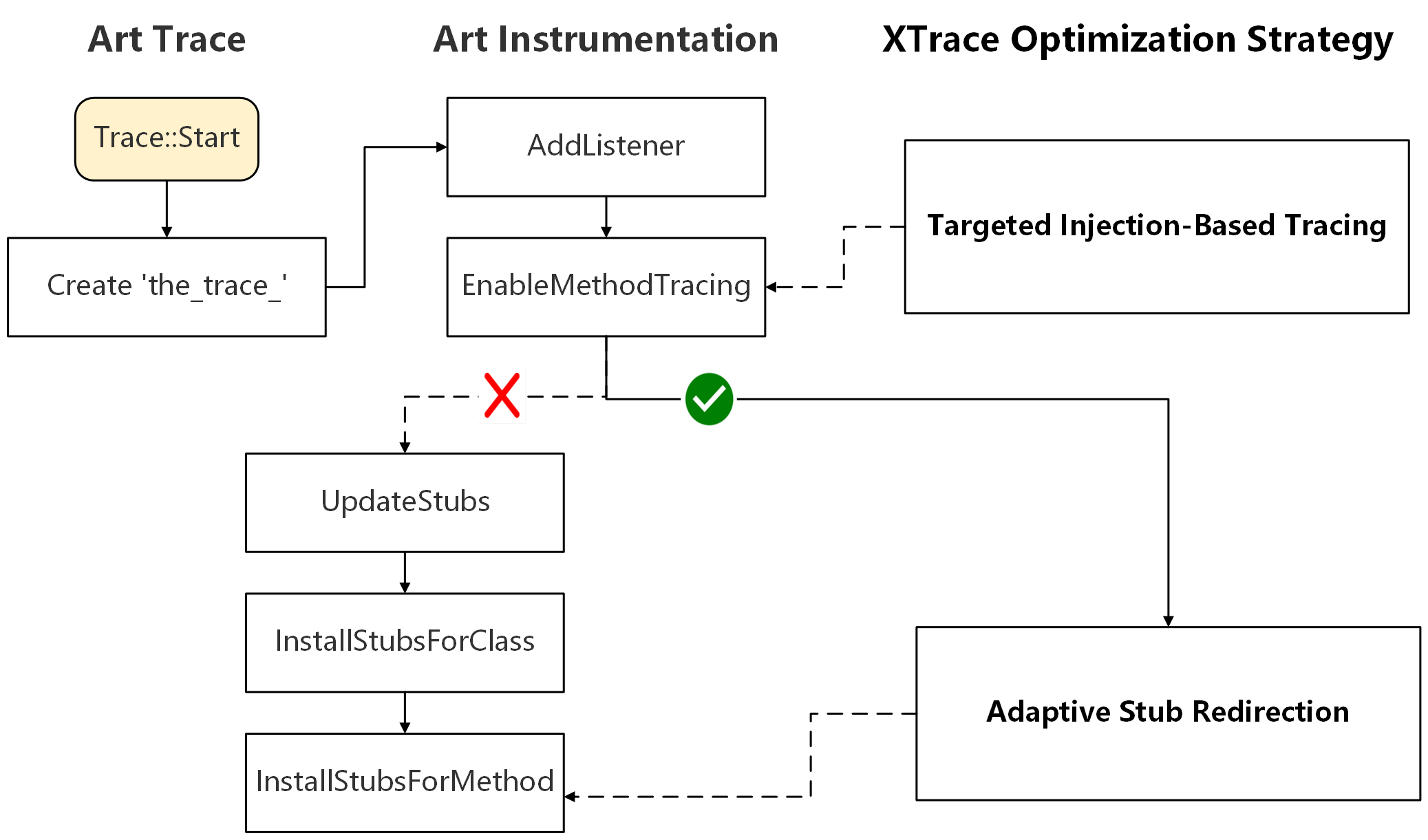}
    \caption{Native vs. XTrace: A high-precision bypass of the global instrumentation flow.}
    \label{fig:xtrace_opt}
\end{figure}

Specifically, we first mount our custom logic onto the native event stream through non-invasive proxying of the system's native trace listener interface, as shown in Figure~\ref{fig:interpreter_execute_code}. This design ensures that XTrace inherits the innate stability and cross-version compatibility of the native instrumentation mechanism. Furthermore, we introduce the following two core optimizations for performance. The result workflow of XTrace is shown in Figure~\ref{fig:xtrace_opt}.

\begin{enumerate}
    \item \textit{targeted injection-based tracing paradigm}: The coarse-grained tracing approach, which employs indiscriminate global instrumentation, introduces substantial performance overhead. To address this limitation, we adopt a dynamic and precise targeting mechanism. As indicated by the red 'X' icon in Figure~\ref{fig:xtrace_opt}, we first hook \textit{Trace::EnableMethodTracing} to a no-op proxy. This effectively disables the native, full-scale instrumentation process of \textit{UpdateStubs}. XTrace then takes over to perform precise, on-demand instrumentation at runtime, targeting exclusively a predefined set of methods. This is accomplished by accurately locating their corresponding \textit{ArtMethod} structures and redirecting their entry points.
    This fundamentally eliminates the massive performance overhead from redundant instrumentation, thereby minimizing the impact of tracing on overall system performance.

    \item \textit{adaptive stub redirection}: The native trace mechanism indiscriminately forces all traced methods into interpretation mode, regardless of whether they have been optimized by JIT or AOT compilation. 
    To address this issue, we propose a strategy called \textit{adaptive stub redirection}. The core principle of this strategy is to first discern the current execution state of a target method before injecting a trace stub. If the method has already been optimized by JIT/AOT compilation, we select a fast-path stub capable of preserving its compiled performance. Conversely, if the method remains in an interpreted state, we utilize a conventional interpreter bridge stub.
    
    This adaptive decision-making is designed to maximally preserve the performance gains from JIT/AOT compilation, avoiding the performance penalty of the native framework's forced mode degradation and thereby achieving a unification of high performance and observability.
\end{enumerate}

\subsubsection{Technical Implementation}
This section delineates the technical blueprint of the XTrace framework, outlining the methodology for transforming a high-overhead native tool into a high-performance, event-driven engine. The exposition proceeds from a high-level architectural overview, which provides a conceptual roadmap, to a detailed examination of the core methodology. 

\begin{figure}
    \centering
    \includegraphics[width=1\linewidth]{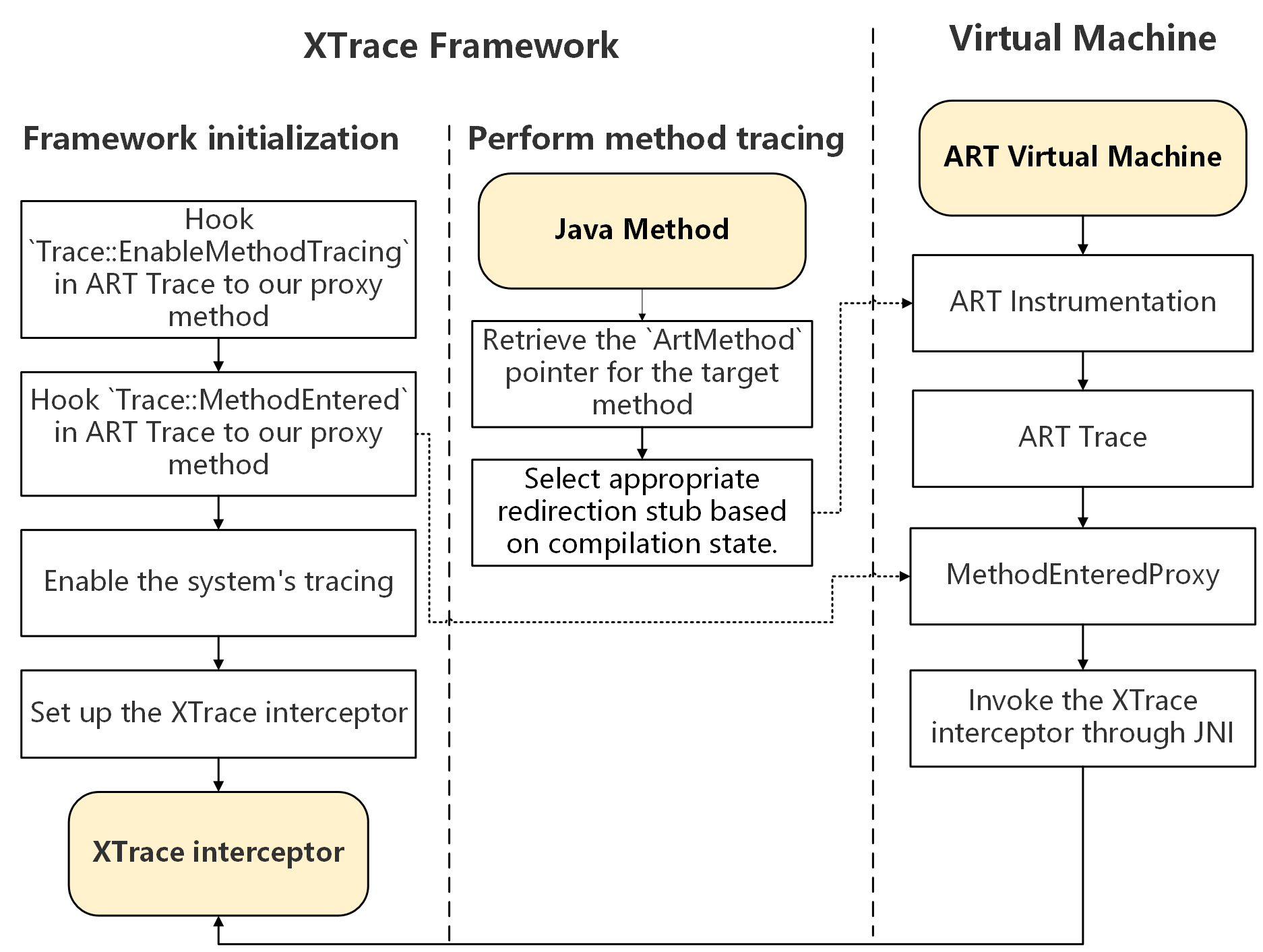}
    \caption{Implementation Architecture of XTrace}
    \label{fig:xtrace_architecture}
\end{figure}

Figure~\ref{fig:xtrace_architecture} illustrates the overall architecture of XTrace, which has two main components: the XTrace Framework and the target ART Virtual Machine. Its core design is a customizable tracing loop that bypasses native global instrumentation through a sophisticated combination of hooking and redirection.

The process starts in the XTrace framework (left). It first hooks the system-wide instrumentation activation function to disable its default behavior and deploy a custom event proxy. When a specific Java method needs to be traced, the framework modifies the method's execution entry point within the virtual machine, redirecting it to ART's instrumentation mechanism.

At runtime, this redirection enables the ART Virtual Machine (right) to trigger an instrumentation event upon the method's execution. This event flows through the ‘ART Trace’ module and is ultimately captured by our deployed event proxy. The proxy then uses a JNI callback to invoke the XTrace Interceptor, returning execution control to the framework for custom trace logic processing.

In essence, this architecture establishes a closed control loop: XTrace intercepts system events, selectively modifies method behavior, and redirects execution flow back to its own customizable logic, completely bypassing the high overhead of the native mechanism.

The core methodology for implementing this architecture is formalized in Algorithm~\ref{alg:our_method}, which details the steps to re-engineer the native tracing mechanism.
\begin{algorithm}[H]
\caption{Adaptive and Precise Method Tracing}
\label{alg:our_method}
\begin{algorithmic}[1]
    \Require Target method set $\mathcal{T}$
    \Statex 
    \State \textbf{Replace} \texttt{EnableMethodTracing} \textbf{with} \texttt{EmptyHandler} \hfill $\triangleright$ \textit{Disable global tracing}
    \For{each method $m \in \mathcal{T}$}
        \State $entry\_point \gets \text{SelectOptimalEntryPoint}(m)$ \hfill $\triangleright$ \textit{Adaptive selection}
        \State \textbf{SetEntryPoint}($m$, $entry\_point$)
    \EndFor
    \State \textbf{SetHandler}(\texttt{MethodEntryEvent}, \texttt{MethodEntryProxy}) \hfill $\triangleright$ \textit{Install dispatcher}
    \State \textbf{ActivateEventEngine}() \hfill $\triangleright$ \textit{Activate event tracing}
    \Statex 
\end{algorithmic}
\end{algorithm}
As formalized in Algorithm~\ref{alg:our_method}, our method proceeds in three key phases:

\emph{Phase 1: suppression of global instrumentation}. The primary performance bottleneck of native tracing is the wholesale replacement of entry points for all methods across all loaded classes, triggered by \textit{EnableMethodTracing}. To eliminate this, we hook \textit{EnableMethodTracing} and replace it with a no-op proxy, \textit{EmptyHandler}(Algorithm~\ref{alg:our_method}, Line 1). This step prevents the default full-scale system-wide instrumentation while keeping the underlying event engine active, thereby enabling our targeted injection in the next phase.
    
\begin{figure}
        \centering
        \includegraphics[width=0.75\linewidth]{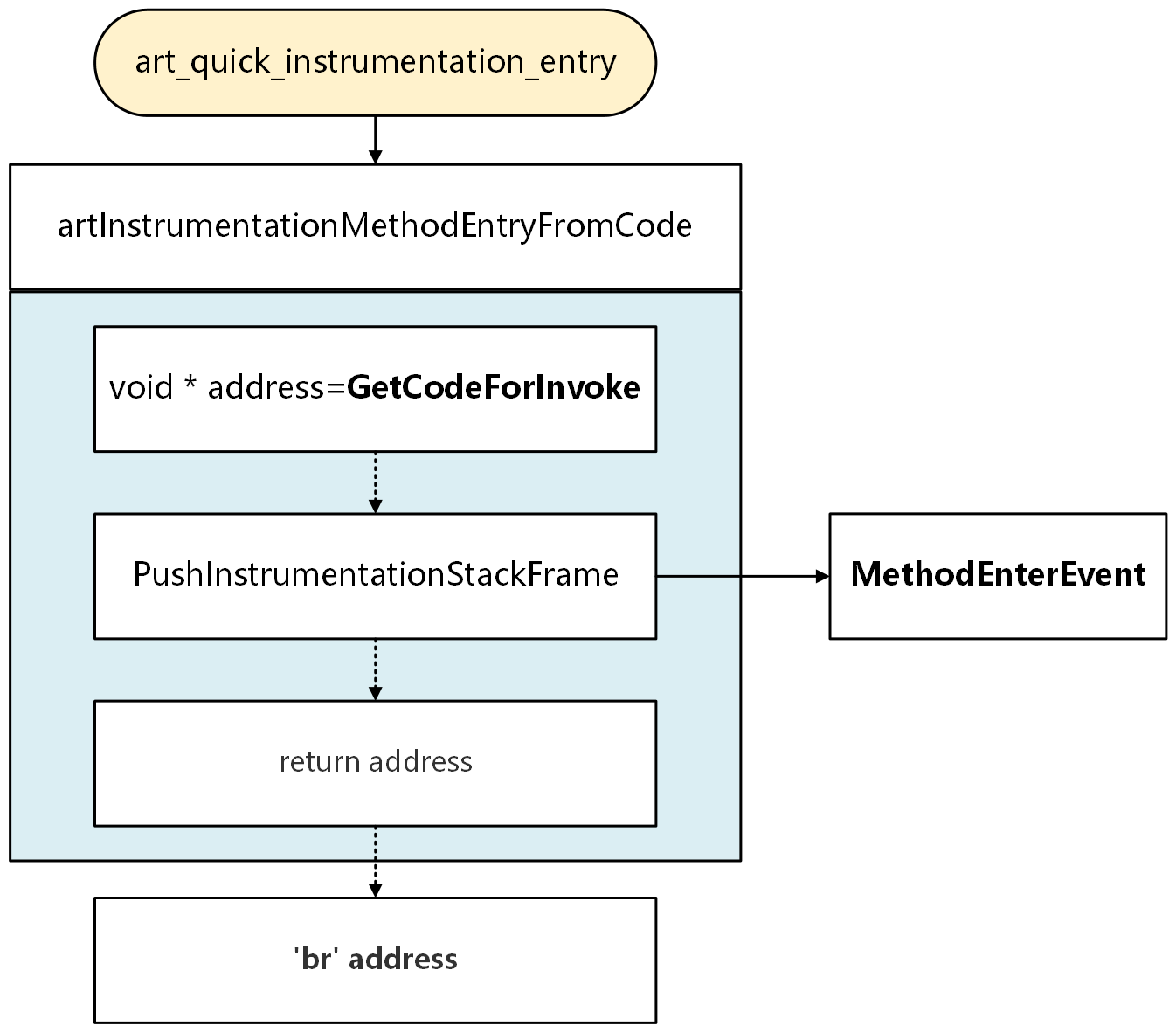}
        \caption{Control Flow of art\_quick\_instrumentation\_entry}
        \label{fig:instrument_code_impl}
    \end{figure}

\emph{Phase 2: targeted injection and proxy deployment}. With global tracing suppressed, the framework modifies only the methods within the target set $\mathcal{T}$ while leaving all other methods untouched (Algorithm~\ref{alg:our_method}, Lines 2-5). This is achieved via two techniques:
\begin{enumerate}[listparindent=\parindent, parsep=0pt]
    \item  Adaptive entry point selection. Unlike the system's default, one-size-fits-all instrumentation, our \textit{SelectOptimalEntryPoint($m$)} function(Line 3) adaptively chooses an optimal stub for each method $m$. Based on its compilation state, it selects either a standard interpreter bridge or, for JIT/AOT-compiled code, the highly performant \textit{art\_quick\_instrumentation\_entry} stub to minimize per-trace overhead.
    
    The superior performance of this stub stems from its inherent context-awareness and restoration capabilities, leveraging ART's native instrumentation support. Its assembly implementation(Figure~\ref{fig:instrumentation_entry_code}) and flowchart(Figure~\ref{fig:instrument_code_impl}) show its core principle: 1) using an internal interface, \textit{GetCodeForInvoke}, to precisely capture the original compiled machine code entry point; 2) executing the trace logic; and 3) performing a direct jump (`br`) back to the original entry point. This ensures that the method seamlessly resumes its original high-performance execution path after tracing.

    \item precise entry point update. The \textit{SetEntryPoint} operation(Line 4) atomically updates the entry address of only the target method $m$. This targeted modification is key to our surgical approach, ensuring no side effects and thus achieving zero collateral overhead for non-target methods themselves.
   \end{enumerate}
   
 After injecting all target methods, \textit{SetHandler}(Line 6) mounts our core dispatcher, \textit{MethodEntryProxy}, onto the system-level \textit{MethodEntryEvent}. This proxy supplants the system's default logic, becoming the central dispatch hub for XTrace at runtime and finalizing its deployment.

    \begin{figure*}
    \begin{lstlisting}[
        language=MyARM,
        frame=single,
        basicstyle=\ttfamily\footnotesize
    ]
    // Instrumentation-related stubs
        .extern artInstrumentationMethodEntryFromCode
    ENTRY art_quick_instrumentation_entry
        SETUP_SAVE_REFS_AND_ARGS_FRAME
        mov   x20, x0          // Preserve method reference in a callee-save.
        mov   x2, xSELF
        mov   x3, sp           // Pass SP
        bl    artInstrumentationMethodEntryFromCode // (Method*, Object*, Thread*, SP)
        mov   xIP0, x0         // x0 = result of call.
        mov   x0, x20          // Reload method reference.
        RESTORE_SAVE_REFS_AND_ARGS_FRAME // Note: will restore xSELF
        REFRESH_MARKING_REGISTER
        cbz   xIP0, 1f         // Deliver the pending exception if method is null.
        adr   xLR, art_quick_instrumentation_exit
        br    xIP0             // Tail-call method with lr set to art_quick_instrumentation_exit.
    1:
        DELIVER_PENDING_EXCEPTION
    END art_quick_instrumentation_entry
    \end{lstlisting}
    \caption{The `art\_quick\_instrumentation\_entry` Instrumentation Stub}
    \label{fig:instrumentation_entry_code},
    \end{figure*}
    
    \emph{Phase 3: activation and runtime dispatching}. With the preparatory phase complete, The final step is to activate the event engine. Our key finding is that this can be achieved by repurposing the native \textit{Debug.startMethodTracingDdms} interface. Invoking it while \textit{EnableMethodTracing} is suppressed fundamentally alters its semantics: its role shifts from initiating a costly, full-scale tracing session to activating the underlying ART event dispatch mechanism in isolation, without triggering any method body replacements.
    
    This lightweight engine then dispatches method entry events to our \textit{MethodEntryProxy}. The proxy efficiently filters these events: it processes the tracing logic only if the current method $m$ is in the target set $\mathcal{T}$ and immediately returns otherwise, effectively ignoring the massive volume of calls to non-target methods with near-zero overhead.

\subsection{Deployment and Operational Workflow}
\begin{figure*}
    \centering
    \includegraphics[width=1\linewidth]{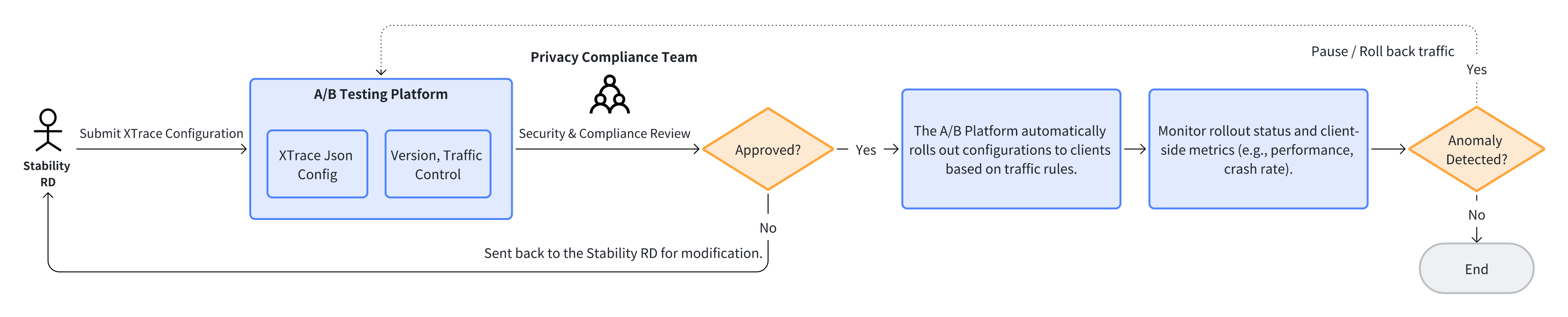}
    \caption{The Release and Risk Control Workflow for XTrace Configurations}
    \label{fig:=xtrace_configuration_deployment}
\end{figure*}

XTrace has been implemented as an SDK (consisting of a native shared object library and a Java wrapper), which is integrated into the applications from Bytedance.
The SDK does not require system privileges, such as root access, as it relies on public Android APIs and operates exclusively within the host applications. We rely on a centralized A/B testing platform to deliever tracing tasks to the SDK, which follows a standardized operational workflow, as illustrated in Figure~\ref{fig:=xtrace_configuration_deployment}.


\begin{figure}[h!]
\begin{lstlisting}[
    language=Java,
    frame=single,
    basicstyle=\ttfamily\footnotesize
]
"dynamic_trace_config": [
  {
    "action": 1, // Action 1: Capture stack trace
    "className": "androidx.window...WindowLayoutComponentImpl",
    "methodName": "addWindowLayoutInfoListener",
    "methodSign": "android.content.Context,androidx...Consumer"
  }
]
\end{lstlisting}
\caption{XTrace Configuration to Diagnose the "Ghost Crash"}
\label{fig:xtrace_configuration} 
\end{figure}

The key stages of the workflow include:
\begin{enumerate}
    \item Submission and versioning: authorized engineers submit configuration drafts through the A/B testing platform. The platform versions all changes, creating a complete and auditable history. Each configuration specifies the target class and method signatures, along with the precise tracing action to be performed, as shown in Figure ~\ref{fig:xtrace_configuration}.
    \item Mandatory compliance review: before activation, each configuration undergoes a mandatory review by a dedicated privacy and security team. This audit ensures the tracing behavior complies with company policies and data protection regulations, such as GDPR.
    \item Canary release: after approval, the configuration is first deployed to a small subset of all active users (typically 0.1\%). This staged rollout minimizes the potential impact of any unforeseen issues.
    \item Real-time monitoring and alerting: during the canary phase, our system continuously monitors key client-side health metrics, like crash and ANR rates. Automated alerts are triggered if these metrics exceed predefined thresholds.
    \item One-click rollback and gradual rollout: if an anomaly is detected, engineers can immediately pause or roll back the configuration. This action instantly stops the delivery of the configuration and invalidates it on devices that have already received it, enabling rapid recovery. Once all key metrics remain stable throughout the canary phase, the configuration is gradually rolled out.
\end{enumerate}


XTrace's dynamic features are designed with data privacy at their core. All dynamic tracing configurations, similar to static tracing, undergo a mandatory compliance review by a Data Protection Officer (DPO). This process enforces data minimization, ensuring that tracing targets only the essential functions required to resolve a specific issue.
On the device, XTrace processes runtime data, like method parameters, exclusively in memory. This data is never written to disk, which significantly reduces the risk of data leakage from local storage.

Once processed, XTrace transfers the data to the company's standard logging pipeline, inheriting its security features. Before upload, an on-device Data Leakage Prevention (DLP) engine sanitizes the data, masking any potential Personally Identifiable Information (PII). The sanitized data is then encrypted at the application layer and transmitted securely over a TLS 1.3 channel.
The backend services only receive and process this anonymized data. As a result, all subsequent analysis for XTrace is performed on aggregated datasets that cannot be traced back to individual users. This aggregated data is subject to a platform-managed 14-day retention policy, after which it is permanently deleted.
\section{Results}
In this chapter, we present and analyze a series of experiments to evaluate the stability, compatibility, performance, and effectiveness of XTrace. We will conduct a detailed evaluation centered around the following three Research Questions (RQs).
\begin{itemize}
    \item RQ1: \textit{Stability and compatibility}: Is XTrace stable and reliable in a large-scale online environment, and is it compatible with mainstream Android system versions?
    \item RQ2: \textit{Performance overhead}: What is the performance overhead introduced by XTrace compared to the baseline and existing mainstream technologies?
    \item RQ3: \textit{Practical effectiveness}: How effective is XTrace in solving real-world, complex engineering problems online?
\end{itemize}
\subsection{RQ1: Stability and Compatibility}
To answer RQ1, we first analyzed the core stability metrics of XTrace from a month-long, large-scale online A/B test covering approximately 108 million daily active users. Table~\ref{tab:stability_metrics} presents a comparison of the data between the treatment group (with XTrace enabled) and the control group (No XTrace, as baseline).
\begin{table*}[htb]
    \centering
    \caption{Impact Analysis of XTrace on Online Stability}
    \label{tab:stability_metrics}
    \renewcommand{\arraystretch}{1.5} 
    \setcellgapes{1pt}
    \makegapedcells
    \renewcommand\theadfont{\bfseries} 
    \renewcommand\theadalign{cc} 
    \begin{tabularx}{\textwidth}{|l|c|c|c|c|c|>{\centering\arraybackslash}X|}
        \hline
        \thead{Metric} & 
        \thead{Control Group \\ ($\approx$54M Users)} & 
        \thead{Experiment Group \\ ($\approx$54M Users)} & 
        \thead{Relative \\ Change} & 
        \thead{95\% Confidence \\ Interval} & 
        \thead{p-Value} & 
        \thead{Non-Inferiority \\ Margin \& Conclusion} \\
        \hline
        Crash User Rate & 0.018490\% & 0.018501\% & +0.0595\% & [-1.191\%, 0.1464\%] & 0.615817 & Pass ($\delta = +0.5\%$) \\
        \hline
        ANR Rate & 0.025480\% & 0.025510\% & +0.1177\% & [-0.2117\%, 0.1265\%] & 0.551651 & Pass ($\delta = +0.8\%$) \\
        \hline
    \end{tabularx}
    \renewcommand{\arraystretch}{1.0} 
\end{table*}

The data in Table~\ref{tab:stability_metrics} shows that the key stability metrics of the XTrace treatment group are nearly identical to the baseline. A statistical test (Z-test) also confirms that there is no significant difference between the two (p > 0.05). This indicates that the introduction of XTrace did not have a negative impact on the application's online stability.

XTrace is engineered for high forward compatibility and low maintenance on Android by relying exclusively on a small, stable, and public subset of the Android Runtime's (ART) native instrumentation interfaces: \textit{UpdateMethodsCode} (to update the target method's entry point), \textit{EnableMethodTracing}, and \textit{MethodEntered}. This design avoids dependencies on internal, volatile ART data structures, significantly mitigating the risk of failures from underlying system changes.

In practice, these native interfaces rarely change. Since Android 5.0, we have only observed minor signature modifications to \textit{MethodEntered} and \textit{EnableMethodTracing} in Android 13 and 14, respectively, while their core functionality remained unchanged. Consequently, adapting XTrace to new Android versions typically only requires updating a symbol mapping table.

To validate this approach, we have evaluated XTrace's compatibility through a 1,200-scenario automated regression test suite on ROMs covering 98\% of our daily active users. The functional correctness was verified with a 99.9\% pass rate.

The maintenance effort for each new Android version, which includes regression testing and adapting to any new interface signatures, is typically just 1-2 person-days. A greater effort would only be necessary in the unlikely event of a fundamental change to the functionality of these underlying ART interfaces—an event we have not yet observed.

\subsection{RQ2: Performance Overhead}
To answer RQ2, we quantified the performance overhead of XTrace from both startup and runtime perspectives and compared it against the baseline and the dynamic instrumentation tool Frida.

The experiments were conducted on a Xiaomi device (Android 15) 
within two cold/hot startup scenarios, with \textit{Log.e} as the tracing target. Each reported value is the average of 10 measurements. Table~\ref{tab:startup_times} shows the measured cold and hot startup times on the host applications.  It indicate that the average startup latency introduced by XTrace is about 6.5ms, far below the user-perceptible threshold, and its overhead is only about 6-10\% of Frida's.

\begin{table}[htb]
  \centering
  \caption{Startup Time Overhead: XTrace vs. Frida (Unit: ms)}
  \label{tab:startup_times}
  \footnotesize
  \begin{tabular}{|wc{1.3cm}|wc{2cm}|wc{1.9cm}|wc{2.1cm}|}
    \hline
    \textbf{Startup Type} & 
    \textbf{\makecell{Baseline-No XTrace \\ (Mean ± Std)}} & 
    \textbf{\makecell{XTrace \\ (Mean ± Std)}} & 
    \textbf{\makecell{Frida \\ (Mean ± Std)}} \\
    \hline
    Cold Startup & 450.2 ± 25.1 & 456.7 ± 25.9 (+6.5ms) & 580.5 ± 45.2 (+130.3ms) \\
    \hline
    Hot Startup  & 180.5 ± 10.3 & 183.9 ± 10.8 (+3.4ms) & 215.8 ± 18.5 (+35.3ms) \\
    \hline
  \end{tabular}
\end{table}

We have also meastured  runtime overhead of XTrace and Frida. Specifically, XTrace's CPU usage overhead is less than 1\% (vs. Frida's \textasciitilde5-8\%), and its per-method call latency is under 0.01ms, which is an order of magnitude lower than Frida's (\textasciitilde0.1-0.2ms).

XTrace's performance advantage stems from its reuse and performance optimization of the system's built-in Instrumentation mechanism. Unlike Frida, which requires memory modification of code segments, instruction translation, and complex trampoline jumps, XTrace merely updates the method entry point, thus avoiding high runtime overhead.

To quantify the impact of XTrace's core mechanisms, we conducted an ablation study comparing the full implementation against two degraded versions:

\begin{itemize}
    \item XTrace-Global: Targeted injection was disabled, forcing global tracing.
    \item XTrace-Interpreter: The adaptive mechanism was disabled, forcing all calls through the interpreter stub.
\end{itemize}

The results, as presented in Table~\ref{tab:ablation_study}, confirm that both mechanisms are critical for performance. Targeted injection is essential for minimizing static overhead. Disabling it (i.e., XTrace-Global) resulted in a prohibitive \textasciitilde64x increase in startup time (from \textasciitilde6.5ms to \textasciitilde418.0ms) and a substantial rise in CPU usage to \textasciitilde37.1\%.

The adaptive stub is crucial for runtime efficiency. Disabling it (i.e., XTrace-Interpreter) had no impact on startup but increased per-call latency by more than 13x (i.e., from <0.01ms to \textasciitilde0.13ms), which is critical when tracing frequently called methods.

In conclusion, both core optimization mechanisms are critical to XTrace's performance, yielding significant advantages over mainstream frameworks like Frida and confirming its viability as a production-grade diagnostic tool.

\begin{table}[htbp]
  \centering
  \caption{Ablation Study of XTrace's Core Mechanisms}
  \label{tab:ablation_study}
  \renewcommand{\arraystretch}{1.2}
   \footnotesize

  \begin{tabular}{|wc{1.7cm}|wc{1.3cm}|wc{2.4cm}|wc{2cm}|}
    \hline
    \textbf{Metrics} &
    \textbf{\makecell{XTrace (Full) \\ \textit{(Baseline)}}} &
    \textbf{\makecell{XTrace-Global \\ (w/o Targeted Injection)}} &
    \textbf{\makecell{XTrace-Interpreter \\ (w/o Adaptive Stub)}} \\
    \hline
    Cold Startup Delay &
    \textasciitilde 6.5ms &
    \makecell{\textasciitilde 418.0ms \\ (\textasciitilde64x increase)} &
    \makecell{\textasciitilde 8.0ms \\ (No impact)} \\
    \hline
    CPU Increase &
    \textless ~1.0\% &
    \makecell{\textasciitilde37.1\% \\ (Significant increase)} &
    \makecell{\textasciitilde1.5\% \\ (Slight increase)} \\
    \hline
    Per-Call Latency &
    \textless 0.01ms &
    \makecell{\textasciitilde 0.03ms \\ (No impact)} &
    \makecell{\textasciitilde0.13ms \\ (\textgreater13x increase)} \\
    \hline
  \end{tabular}
  \renewcommand{\arraystretch}{1.0}
\end{table}




\subsection{RQ3: Practical Effectiveness}
To answer RQ3, we systematically evaluate XTrace's practical effectiveness from two distinct problem domains: stability and performance optimization. As described in the Section~\ref{sec:overview}, XTrace has already demonstrated its pivotal role in the stability domain by locating and fixing a critical online crash (affecting over 40,000 users daily). Therefore, to further demonstrate its breadth of applicability, this section focuses on the latter, illustrating its capabilities in UI performance optimization through an in-depth case study.

     \textit{Problem background \& limitations of existing tools}: Users once reported significant jank while scrolling through a video feed in an app from ByteDance. We first used Perfetto~\cite{perfetto}, an industry-leading system-wide tracing tool, for analysis. Although Perfetto's timeline clearly revealed that \textit{Choreographer.doFrame} frequently exceeded the 16.6ms budget and that the bottleneck was in the \textit{measure} phase (one of the three phases in \textit{View} drawing). This exposed its fundamental limitations in precise attribution. These limitations stem from its design: first, it relies on static, pre-instrumented tracepoints (e.g., `ATRACE\_BEGIN`) and cannot dynamically trace arbitrary Java methods, causing the call chain to break at the application logic layer. Second, it cannot automatically correlate high-level events (like the start of a frame) with low-level function calls (like a specific \textit{View} instance's \textit{measure} call), nor can it quantify the call frequency of a particular instance. Consequently, while Perfetto could confirm the symptom, it could not pinpoint the illness, that is, which specific component in the complex view hierarchy was causing the unnecessary redraw overhead and for what reason.
     
     \textit{XTrace methodology \& key findings}: To overcome these limitations, we deployed XTrace. Leveraging its ART instrumentation-based dynamic tracing capabilities, we simultaneously targeted and traced both the high-level frame-drawing event \textit{FrameDisplayEventListener.run} and the low-level layout calculation function \textit{View.measure}. XTrace's results provided two key insights unattainable with traditional tools:
    \begin{enumerate}
    \item  XTrace captured and presented the complete Java call stack for every \textit{measure} call that occurred within the lifecycle of a \textit{FrameDisplayEventListener.run} event. This revealed an unbroken causal path from the top-level business logic down to the multiple low-level \textit{measure} calls, bridging the analytical gap left by traditional tools.
    \item  It further precisely attributed these redundant calls to the same \textit{RelativeLayout} instance and quantified its call frequency as 3 times per frame.
    \end{enumerate}
    These two insights collectively pointed to multiple redraws of a specific component as the root cause.

With this decisive evidence from XTrace, which included the complete causal chain and instance information, the development team quickly identified a layout constraint conflict and fixed it within half a day. In the subsequent A/B testing, the treatment group's UI frame drop rate decreased by a significant 25.95\%, with a tangible improvement in scrolling smoothness.

In summary, these two cases (including the one from Section~\ref{sec:overview}) collectively demonstrate that XTrace is highly effective in practice at diagnosing and resolving complex multithreading and performance issues. Its technical advantages translate directly into measurable business and product value. By providing decisive evidence—such as complete causal chains and runtime context, which are unattainable with conventional tools—XTrace reduces the mean time to diagnosis (MTTD) for complex online issues from days to mere hours, an efficiency improvement of over 90\%. This has not only fixed critical crashes affecting tens of thousands of users but also reduced the UI frame drop rate in core scenarios by 25.95\%, proving its immense potential in ensuring service quality and enhancing user experience.
\section{Related Work} 
Dynamic analysis on mobile platforms is a critical yet challenging field. Existing technical approaches can be broadly categorized into three types: intrusive dynamic instrumentation frameworks, system-level tracing tools, and the utilization of the virtual machine's native interfaces. Our work, XTrace, is situated within this technical landscape: its design philosophy originates from the third category but fundamentally overcomes its inherent limitations.
\subsection{Intrusive Dynamic Instrumentation Frameworks}
A predominant approach to dynamic analysis involves intrusive runtime hooking, with Frida~\cite{Frida} and the Xposed~\cite{Xposed_Framework} Framework being prominent examples. These tools typically operate by directly modifying the method body of a target function at runtime, inserting a "trampoline" code to redirect the execution flow.
While these frameworks offer powerful flexibility, their "Intrusive Intervention" model introduces inherent risks to stability and compatibility. They rely on modifying undocumented and volatile internal structures of the virtual machine (e.g., the \textit{ArtMethod} struct), making them fragile in the face of Android's high fragmentation, diverse manufacturer ROMs, and frequent version iterations. This vulnerability is well-documented in their public issue trackers, which contain numerous crash reports related to compatibility. In contrast, XTrace's design philosophy fundamentally avoids such direct memory manipulation of method implementations, aiming for a more stable solution.
\subsection{ System-level and Kernel-level Tracing Tools}
Another class of tools operate at the system or kernel level, such as Perfetto~\cite{perfetto} and its predecessor, Systrace~\cite{Systrace_Doc}. These tools leverage statically compiled tracepoints within the Android framework and the Linux kernel (e.g., ftrace) to provide a low-overhead, system-wide view of performance events.
However, this reliance on static tracepoints is also their fundamental limitation. First, their coverage is strictly limited to the tracepoints pre-placed by system and framework developers, making them unable to automatically trace arbitrary, non-instrumented application-level Java methods. Second, they lack the flexibility to add new tracing logic on-demand at runtime.
Consequently, developers often face a dilemma: they can observe high-level system symptoms (e.g., a long frame rendering time) but lack the fine-grained, application-level context to pinpoint the root cause (e.g., which specific piece of business logic triggered a redundant layout calculation). Ultimately, when the execution flow transitions from the system framework into application business logic, the call stacks provided by these tools lose their fine-grained continuity. As they can only recognize a few manually inserted tracepoints by the developer, vast "tracing blind spots" exist within critical business logic, making it impossible for developers to construct a complete causal chain from system symptom to code-level root cause.
In response, XTrace enables the dynamic tracing of any Java method at the ART virtual machine level and supports flexible, custom logic injection, thereby directly providing the complete, application-aware call stacks necessary for root cause analysis.
\subsection{ART's Native Instrumentation and Debugging Interface}
The Android RunTime (ART) provides a native Instrumentation interface derived from the Java Virtual Machine Tool Interface (JVMTI)~\cite{JVMTI_Spec}\cite{jvmti_example}, primarily designed for debuggers and profilers. This interface exposes a mechanism to listen for events such as method entry and exit, providing a theoretical foundation for building non-invasive tracing tools.
However, the default implementation of this interface is widely considered unsuitable for production environments by both academia and industry due to its inherent performance deficiencies. Its main problems are twofold: first, indiscriminate global injection, which introduces significant performance overhead; and second, forced degradation to interpreter execution, which causes severe performance degradation. Prior to our work, these well-known performance bottlenecks rendered this technical path largely infeasible for large-scale deployment in production environments.
The core contribution of XTrace is precisely in systematically addressing the aforementioned performance deficiencies of the native interface, thereby transforming this overlooked technical route into an efficient, production-ready solution.
\section{Conclusions}

Unlike mainstream frameworks like Frida that use "intrusive intervention"—patching volatile, undocumented system internals and thus suffering from instability across different Android versions and devices, XTrace ensures stability by exclusively using the Android Runtime's (ART) official, stable instrumentation interfaces, thereby inheriting the native system's high compatibility.

However, these native interfaces are too slow for production use on their own. XTrace's core contribution lies in overcoming this bottleneck with two key innovations: (1) Targeted Injection: Minimizes static overhead by activating tracing only for specific, required methods.
(2) Adaptive Execution: Reduces runtime latency by dynamically switching between a fast compiled stub and a full-featured interpreter stub. These optimizations transform the stable but slow native mechanism into a high-performance, production-ready solution.

 XTrace delivers a solution that is simultaneously stable, performant, and high-coverage framework that significantly improves diagnostic efficiency for complex online issues, accelerating development cycles and enhancing application stability.
It shows that it is possible to achieve comprehensive runtime observability without sacrificing stability or performance, offering a new path for future research in mobile system dynamic analysis.

\bibliographystyle{ACM-Reference-Format} 
\bibliography{ref} 

\end{document}